\documentclass[onecollarge]{svjour2}       
\smartqed  
\usepackage{graphicx}
\usepackage{amsmath, amssymb}
\usepackage{amsfonts}
\usepackage{multicol}

\usepackage{ascmac}
\usepackage{fancybox}
\usepackage{type1cm}
\usepackage{comment}

\usepackage{cite}

\newcommand{\todayd}{\the\year/\the\month/\the\day}
\newcommand{\eq}[1]{\begin{equation} #1 \end{equation}}
\newcommand{\eqa}[2]{\begin{equation} #1 \label{#2} \end{equation}}
\newcommand{\del}{\partial}

\newcommand{\la}{\langle}
\newcommand{\ra}{\rangle}
\newcommand{\dis}{\displaystyle}

\newcommand{\bib}{\bibitem}
\newcommand{\appnum}[1]{\renewcommand{\theequation}{#1.\arabic{equation}}
\setcounter{equation}{0}}

\newcommand{\mx}[1]{\begin{pmatrix}#1\end{pmatrix}}

\def \Im{{\rm Im\,}}

\newcommand{\bs}{\boldsymbol}

\def\rnum#1{\resizebox{0.5em}{\height}{\expandafter{\romannumeral #1}}}
\def\Rnum#1{\resizebox{0.5em}{\height}{\uppercase\expandafter{\romannumeral #1}}}

\begin{document}

\title{Anomalous System Size Dependence of Large Deviation Functions for Local Empirical Measure}


\author{Naoto Shiraishi}


\institute{Naoto Shiraishi \at
              Department of Basic Science, The University of Tokyo, Tokyo, 153-8902, Japan \\
             \email{shiraishi@noneq.c.u-toyko.ac.jp}           
          }

\date{Received: date / Accepted: date}

\maketitle

\begin{abstract}

We study the large deviation function for the empirical measure (the time-averaged density) of diffusing particles at one fixed position.
We find that the large deviation function exhibits anomalous system size dependence in systems with translational symmetry if and only if they satisfy the following conditions: (i) there exists no macroscopic flow, and  (ii) their space dimension is one or two.
We investigate this anomaly by using a contraction principle.
We also analyze the relation between this anomaly and the so-called long-time tail behavior on the basis of phenomenological arguments.

\end{abstract}

\section{Introduction}
\label{intro}

The large deviation theory has been intensively studied in non-equilibrium statistical physics~\cite{SSFT, Derrida, Bertini, em, eC, glass, photon, earthquake, motor, NS, meso, math}.
In particular, since the discovery of the fluctuation theorem, which claims a symmetric property of the large deviation function for the time-averaged entropy production~\cite{SSFT}, large deviation functions for various time-averaged quantities have been investigated.
These include the additivity principle in non-equilibrium steady states~\cite{Derrida, Bertini}, some relations in non-equilibrium thermodynamics~\cite{em, eC}, the glass transition~\cite{glass}, the photon emission~\cite{photon}, frequency of earthquakes~\cite{earthquake}, and an operational method for calculating the time-averaged current~\cite{NS}.
As stated above, the importance of the large deviation functions for time-averaged quantities has been increasingly recognized.

Here, let us consider the scenario of observing time-averaged quantities in laboratory experiments.
In many cases, we measure such quantities at one fixed position because measurements at one position is much easier than those over a large area.
Therefore, many studies on large deviation functions for such time-averaged ``local quantities" arise in various contexts such as the molecular motors in living organisms~\cite{motor}, transports in mesoscopic systems~\cite{meso}, and stochastic processes in applied mathematics~\cite{math}.
With the above mentioned background, in this paper, we study a new aspect of the large deviation function for time-averaged local quantities.

Specifically, we investigate the large deviation function for the occupation frequency at one fixed position, which we refer to as a {\it local empirical measure}, in systems with translational symmetry.
The local empirical measure at a position ${\bs x}\in {\mathbb R}^d$ is defined as 
\eq{\mu _{\bs x}(\tau ):=\frac{1}{\tau}\int_0^\tau \rho _{\bs x}(t)dt,}
where $\rho _{\bs x}(t)$ is a fluctuating density at ${\bs x}$ and at time $t$, and $d$ is the space dimension.
The large deviation function for $\mu _{\bs x}(\tau )$ is denoted by $I(\mu )$. That is, 
\eqa{\mathrm{Prob}(\mu _{\bs x}(\tau )=\mu )\sim e^{-\tau I(\mu )}.}{LDF}
Here, ``$A\sim B$" indicates $\lim_{\tau \to \infty}\ln A/\tau =\lim_{\tau \to \infty}\ln B/\tau$.
The main discovery of this paper is the anomalous system size dependence of $I(\mu )$.

Naively, we expect that $I(\mu )$ does not depend on the system size in the thermodynamic limit because $I(\mu )$ is a local quantity.
For systems in more than three dimensions, such a naive expectation holds true.
Surprisingly, however, the naive expectation fails for some systems in one and two dimensions.
Concretely, in the thermodynamic limit, the variance of $\mu _{\bs x}(\tau )$, which can be calculated from $I(\mu )$, diverges in proportion to system size $L$ in one dimension, and $\ln L$ in two dimensions.

\begingroup
\renewcommand{\arraystretch}{1.5}
\begin{table}
\caption{Anomaly in the variance of $\mu$ in the long-time limit and the thermodynamic limit.}
\label{table}       
\begin{tabular}{c||l|l}
\hline\noalign{\smallskip}
dimensions & taking the long-time limit first & taking the thermodynamic limit first \\ \hline
one & $=\Theta (L)$ & $=\Theta \left({1}/{\sqrt{\tau}}\right)$ \\
two & $=\Theta (\ln L)$ & $=\Theta \left({\ln \tau}/{\tau}\right)$ \\
three and more &  $=\Theta (1)$ (standard dependence) & $=\Theta \left({1}/{\tau}\right)$ (standard dependence) \\
\noalign{\smallskip}\hline
\end{tabular}
\end{table}
\endgroup

This anomaly is also understood as the problem that the order of the long-time limit and the thermodynamic limit cannot be exchanged (see Table \ref{table}).
If we fix $\tau$ and take the thermodynamic limit $L\to \infty$ first, $\mathrm{Prob}(\mu _{\bs x}(\tau )=\mu )$ does not satisfy the formula (\ref{LDF}), which is the definition of the large deviation function.
Specifically, the variance of $\mu _{\bs x}(\tau )$ decays as $\Theta ({1}/{\sqrt{\tau }})$ in one dimension and as $\Theta ({\ln \tau }/{\tau })$ in two dimensions for $\tau \to \infty$, which is slower than $\Theta (1/\tau )$ observed in standard cases.
Here, $\Theta$ is defined as follows: $f(x)=\Theta (g(x))$ indicates that there exist positive real numbers $a$ and $b$ such that $a\cdot g(x)\leq f(x)\leq b\cdot g(x)$ for any $x$ larger than some number.
Although the existence of the anomalous time dependence is already known in some specific models such as random walks~\cite{RW} and symmetric simple exclusion process (SSEP)~\cite{otSSEP}, its universal features have not been demonstrated.
In this paper, we present general arguments for the singularity in connection with the anomalous system size dependence.

This paper is organized as follows.
In Sec. I\hspace{-.1em}I, for the purpose of grasping properties of the anomaly, we analyze a solvable microscopic model, specifically independent random walks on a lattice with discrete translational symmetry.
By solving this model, we suggest the essential conditions for this anomaly.
In Sec. I\hspace{-.1em}I\hspace{-.1em}I, for systems whose distribution functions follow the Fokker-Planck equation, we derive this anomaly by using a contraction principle.
In Sec. I\hspace{-.1em}V, from a phenomenological viewpoint, we analyze the relation between this anomaly and the long-time tail behavior~\cite{Alder, ltt, ltt2, ltt3, ssltt, lttstress}.

\section{Microscopic model - a solvable example}

In this section, in order to confirm the existence of the anomaly, we study a simple solvable example.
Consider a $d$-dimensional cubic lattice with a periodic boundary condition. 
The length of the lattice in direction $r\in \{x, y, z, \cdots \}$ is $L_{r}$.
The site in the lattice is denoted by ${\bs i}=\{i_r\}\in {\mathbb N}^d$, where $0\leq i_{r}<L_{r}$.
Suppose that there are $N$ particles on the lattice.
A position of the $s$-th particle at time $t$ is denoted by $\bs{x}_s(t)$.

The particles do not interact with each other.
The time evolution of each particle is described by a discrete-time Markov chain with a transition matrix $S$, where
\eq{S_{{\bs j}, {\bs i}}=\mathrm{Prob}\left({\bs x}_s(t+1)={\bs j}|{\bs x}_s(t)={\bs i}\right).}
The migration length of a particle in one transition is bounded by a finite value independent of $L_{r}$. 
Namely, by setting $\alpha _r \equiv i_r-j_r\mod L_r$ with $-L_r/2\leq \alpha _r <L_r/2$, a necessary condition for $S_{{\bs j}, {\bs i}}\neq 0$ is that $|\alpha _r|$ is less than a given finite value independent of $L_{r}$ in each direction ${r}$.
We also assume that the transition matrix $S$ is periodic with a period of $M_{r}$ in each direction ${r}$.
For example, when $d=2$ and the periods of transition matrix in $x$ and $y$ direction are $M_x$ and $M_y$ (and there are integers $l_x$ and $l_y$, which satisfy $M_xl_x=L_x$ and $M_yl_y=L_y$), $S_{{\bs j},{\bs i}}=S_{{\bs j'}, {\bs i'}}$ holds if ${\bs i}-{\bs j}={\bs i'}-{\bs j'}$ and $i_{x}\equiv i'_{x}\mod M_{x},\ i_{y}\equiv i'_{y}\mod M_{y}$.

We denote by $P({\bs x}(0))$ the initial probability distribution.
Now, we define the local empirical measure at the origin ${\boldsymbol{0}}$ as
\eq{\mu _{\boldsymbol{0}}(\tau):=\frac{1}{\tau}\sum_{s=1}^{N}\sum_{t=0}^{\tau}\delta _{{\boldsymbol{0}}, {\bs x}_s(t)},}
where $\delta$ is the Kronecker delta.
First, we take the long-time limit $\tau \to \infty$.
Then, we take $N$ and $l_{r}$ to be sufficiently large under the condition that the particle density $\rho :=N/\prod_{r}L_{r}$, $M_{r}$, and $S_{{\bs j}, {\bs i}} (0\leq i_{r}\leq M_{r})$ are fixed.
In this setting, we analyze the system size dependence of the variance of $\mu _{\boldsymbol{0}}(\tau)$.

By using the fact that particles move independently, we derive the variance of $\mu _{\boldsymbol{0}}(\tau)$ as follows. 
We denote the local empirical measure of one particle by $\nu _{\boldsymbol{0}}(\tau ):={1}/{\tau}\sum_{t=0}^{\tau}\delta _{{\boldsymbol{0}}, \bs{x}_1(t)}$.
Calculating the variance of $\nu _{\boldsymbol{0}}(\tau )$ and multiplying this value by $N=\prod_{r}L_{r}\rho$, we obtain the variance of $\mu _{\boldsymbol{0}}(\tau)$.

\subsection{Systems in one dimension}

Let $M$ be the period of the transition matrix $S$, and $L$ ($=Ml$) be the length of the system.
We define a transfer matrix $A(h)$ as
\eqa{A_{j,i}(h):=e^{h\delta _{0,j}}S_{j, i},}{defA}
where $0\leq i,j\leq L-1$ and $h$ is a real number.
We denote by $[x(t)]_{t=0}^{\tau}$ a path for the particle $1$ from time $t=0$ to $t=\tau$.
We also denote by $P([x(t)]_{t=0}^{\tau})$ the probability for paths with initial probability distribution $P(x(0))$.

We define $\left<\right>$ as an ensemble average of trajectories generated by the transition matrix $S$.
Then, $\left<e^{h\tau \nu _{0}(\tau )}\right>$ satisfies 
\begin{align}
\left<e^{h\tau \nu _{0}(\tau )}\right>&:=\sum_{[x(t)]_{t=0}^{\tau}}e^{h\tau \nu _{0}(\tau )}P([x(t)]_{t=0}^{\tau})  \notag \\
&=\sum_{[x(t)]_{t=0}^{\tau}}e^{h\tau \nu _{0}(\tau )}\prod_{t=0}^{t={\tau}-1}S_{x(t+1), x(t)}P(x(0)) \notag \\
&={\bs e}A(h)^{\tau}{\bs P}_0. \label{hop}
 \end{align}
$\sum_{[x(t)]_{t=0}^{\tau}}$ represents the summation over all possible paths. 
$\bs{P}_0$ is a vector representation of $P(x(0))$.
${\bs e}$ indicates an $L$-dimensional vector $(1,1,1\cdots ,1)$.
Using the transfer matrix method, we obtain
\eqa{\lim_{{\tau} \to \infty}\frac{1}{{\tau}}\log \left({\bs e}A(h)^{\tau}{\bs P}_0\right)=\log \Lambda _{\mathrm max}(h),}{ortho}
where $\Lambda _{\mathrm max}(h)$ represents the maximum eigenvalue of the matrix $A(h)$.
Note that ${\bs \phi }_{\mathrm max}(h)$, the corresponding right eigenvector to $\Lambda _{\mathrm max}(h)$, and ${\bs P}_0$ are not orthogonal.
It is because ${\bs P}_0$ is a nonnegative vector and ${\bs \phi }_{\mathrm max}(h)$ is a positive vector, which is implied by the Perron-Frobenius theorem.
It follows from Eqs. (\ref{hop}) and (\ref{ortho}) that
\begin{equation}\lim_{{\tau}\to \infty}\tau ^{n-1}\la \nu _{0}(\tau )^n\ra _{\mathrm c}
=\left.\left(\frac{\partial}{\partial h}\right)^n\log \Lambda _{\mathrm max}(h)\right|_{h=0}, \label{cumh} \end{equation}
where $\la \ra _{\mathrm c}$ represents cumulants.
We define cumulants as $\sum_{n=0}^{\infty}h^n/n!\cdot \la x^n\ra =\exp \left(\sum_{n=0}^{\infty}h^n/n!\cdot \la x^n\ra _{\mathrm c}\right)$.
The first cumulant $\la \nu _{0}(\tau )\ra _{\mathrm c}$ is equal to the expectation value, and the second cumulant $\la \nu _{0}(\tau )^2\ra _{\mathrm c}$ is equal to variance.

Now, our goal is to derive $\Lambda _{\mathrm max}(h)$.
The matrix $A(h)$ is written as $A(h)=S+\left(h+h^2/2\right)X+O(h^3)$, where $X$ is defined as
\begin{equation}
X:=\left(
   \begin{array}{cccc}
      S_{0,0} & S_{0,1} & \ldots & S_{0,Ml-1} \\
      0 & 0 & \ldots & 0 \\
      \vdots & \vdots & \ddots & \vdots \\
       0 & 0 & \ldots & 0 
    \end{array}
  \right).
\end{equation}
Let ${\bs v}\in \mathbb{Q}^L$ be the right eigenvectors of $S$.
It follows from Bloch's theorem~\cite{Brillouin} that $\bs{v}$ is written as
\begin{equation}\left( \begin{array}{c}
k_0\\
k_1\\
\vdots \\
k_{M-1} \\
k_0e^{i\frac{1}{l}\theta} \\
\vdots \\
k_{M-1}e^{i\frac{1}{l}\theta} \\
k_0e^{i\frac{2}{l}\theta} \\
\vdots \\
k_{M-1}e^{i\frac{l-1}{l}\theta} \\
\end{array} \right), \label{veck}
\end{equation}
where $\theta =2\pi b$, $b=-{l}/{2}+1, -{l}/{2}+2\cdots {l}/{2}$ when $l$ is even, and $b=-{(l-1)}/{2}, -{(l-1)}/{2}+1\cdots {(l-1)}/{2}$ when $l$ is odd.
$k_0 \cdots k_{M-1}$ are coefficients.
Note that $(k_0 \cdots k_{M-1})^{\top}$ is an eigenvector of a matrix $S'\in \mathbb{R}^{M\times M}$, where $S'_{j,i}:=\sum_{x=0}^{l-1}\sum_{y=0}^{l-1}S_{j+xM, i+yM}$ and $\top$ represents a transposition.
We can easily check that (\ref{veck}) are eigenvectors of $S$ and that (\ref{veck}) takes $M\times l$ kinds of different vectors.
Thus, all eigenvectors of $S$ is written as (\ref{veck}).
Let $\lambda _{b, m}$ be the $m$-th largest eigenvalue of $S$ when $\theta =2\pi b$, and ${\bs v}_{b, m}$ and ${\bs u}_{b, m}$ be the corresponding right and left eigenvectors.
${\bs v}_{b, m}$ and ${\bs u}_{b, m}$ are normalized as ${\bs u}_{b, m}{\bs v}_{b, m}=1$.
We denote the value of $k_j$ in ${\bs v}_{b, m}$ as $k_j(b, m)$.
The maximum eigenvalue of $S$ is $\lambda _{0, 0}=1$.
Thus, $\Lambda _{\mathrm max}(h)$ and ${\bs \phi }_{\mathrm max}(h)$ are expanded in $h$ as
\begin{align}
\Lambda _{\mathrm max}(h)&=1+\left(h+\frac{h^2}{2}\right)a_1+\left(h+\frac{h^2}{2}\right)^2a_2+\cdots ,\\
{\bs \phi }_{\mathrm max}(h)&={\bs v}_{0, 0}+\left(h+\frac{h^2}{2}\right)\sum_{\substack{{b, m}\\{(b, m)\neq (0, 0)}}}p_{b, m}{\bs v}_{b, m}+\left(h+\frac{h^2}{2}\right)^2\sum_{b, m}q_{b, m}{\bs v}_{b, m}+\cdots ,
\end{align}
where $a_1$, $a_2$, $p_{b, m}$, and $q_{b, m}$ are coefficients.
By setting $n=2$ in Eq. (\ref{cumh}), the variance of $ \nu _{0}(\tau )$ is expressed as
\eqa{\lim_{\tau \to \infty}\tau \left \langle \nu _{0}(\tau )^2\right \rangle _c=2a_2+a_1-a_1^2.}{Vc}
Here, $a_1$ and $a_2$ are written as
\begin{align}
a_1&={\bs u}_{0, 0}X{\bs v}_{0, 0}=\frac{1}{Ml}\sum_{i=0}^{l-1}\sum_{j=0}^{M-1}k_j(0, 0)S_{0,Mi+j}, \\
a_2&=\frac{1}{Ml}\sum_{\substack{{b, m}\\{(b, m)\neq (0, 0)}}}\frac{{\bs u}_{0, 0}X{\bs v}_{b, m}}{\lambda _{0, 0}-\lambda _{b, m}} \lambda _{b, m}=\frac{1}{Ml}\sum_{\substack{{b, m}\\{(b, m)\neq (0, 0)}}}\frac{{\bs u}_{0, 0}X{\bs v}_{b, m}}{1-\lambda _{b, m}} \lambda _{b, m}.
\end{align}
${\bs u}_{0, 0}X{\bs v}_{b, m}$ is in proportion to ${1}/{Ml}$.
Note that $a_1=\Theta (1/l)$.
When $l$ is sufficiently large, we obtain
\eqa{2a_2\simeq B\left(\frac{1}{Ml}\right)^2\sum_{\substack{{b, m}\\{(b, m)\neq (0, 0)}}}\frac{\lambda _{b, m}}{1-\lambda _{b, m}},}{prea2}
where $B$ is a constant independent of $l$.
The dominant contribution to the right hand side in (\ref{prea2}) is by terms with large $\lambda _{b, m}/\left({1-\lambda _{b, m}}\right) $.
For large $l$, the eigenvalue $\lambda _{b, m}$ takes a value near $1$ only when $m=0$ and $\left|{b}/{l}\right|\ll 1$.
Therefore, abbreviating $\lambda _{b, 0}$ to $\lambda _b$, we can rewrite the right hand side in (\ref{prea2}) as 
\eqa{2a_2\simeq B\left(\frac{1}{Ml}\right)^2\sum_{b\neq 0}\frac{\lambda _b}{1-\lambda _b}.}{a2}

Here, because we can reselect a larger $M$ if necessary, without loss of generality, we assume that particles cannot move more than $M$ sites in one transition.
Then, we define a matrix 
\eqa{S_b:=\left(
    \begin{array}{cccc}
      S_{0,0} & S_{0,1}+S_{M,1}e^{-i\frac{2\pi b}{l}} & \ldots & S_{0,M-1}+S_{M,M-1}e^{-i\frac{2\pi b}{l}} \\
      S_{1,0}+S_{1,M}e^{i\frac{2\pi b}{l}} & S_{1,1} & \ldots & S_{1,M-1}+S_{M+1,M-1}e^{-i\frac{2\pi b}{l}} \\
      \vdots & \vdots & \ddots & \vdots \\
       S_{M-1,0}+S_{M-1,M}e^{i\frac{2\pi b}{l}} & S_{M-1,1}+S_{M-1,M+1}e^{i\frac{2\pi b}{l}} & \ldots & S_{M-1,M-1}
    \end{array}
  \right).}{Sm}
The maximum eigenvalue of $S_b$ is $\lambda _b$, and let the corresponding right and left eigenvector be $\bs{\psi}_b$ and $\bs{\xi}_b$.
$\bs{\psi}_b$ satisfies $\bs{ \psi}_b:=\left(k_0(b, 0), \cdots ,k_{M-1}(b, 0)\right)^{\top}$.
When $\left|{b}/{l}\right|\ll 1$, $S_b$ is expanded as 
\eq{S_b\simeq S_0+\left(i\frac{2\pi b}{l}-\frac{1}{2}\left(\frac{2\pi b}{l}\right)^2\right)Z+O\left(\left(\frac{2\pi b}{l}\right)^3\right),}
where $Z$ is given by
\eq{Z:=\left(
    \begin{array}{cccc}
      0 & -S_{M,1} & \ldots & -S_{M,M-1} \\
      S_{1,M} & 0 & \ldots & -S_{M+1,M-1} \\
      \vdots & \vdots & \ddots & \vdots \\
      S_{M-1,M} & S_{M-1,M+1} & \ldots & 0
    \end{array}
  \right).}
Here, $Z$ corresponds to an operator that gives a particle current passing from $x_{M}$ to $x_{M-1}$.
$\lambda _b$ is expanded as
\eq{\lambda _b=1+i\frac{2\pi b}{l}\cdot J-\left(\frac{2\pi b}{l}\right)^2\cdot C+\cdots .}
Here, $C$ is a quadratic coefficient, which depends only on $Z$ and $S_{0}$.
$J$ is defined as $J:=\bs{\xi}_0Z\bs{\psi}_0$ and $J$ indicates particle flow in steady states.

Eq. (\ref{a2}) takes qualitatively different values depending on whether $J=0$ or $J\neq 0$.
For $J=0$, which means that there exists no macroscopic flow, the right hand side in (\ref{a2}) is
\eq{\left(\frac{1}{Ml}\right)^2\sum_{b\neq 0}\frac{\lambda _b}{1-\lambda _b}\simeq \left(\frac{1}{Ml}\right)^2\sum_{b\neq 0}\frac{l^2}{4\pi ^2b^2\cdot C}\simeq D}
for large $l$, where $D$ is a constant independent of $l$.
Here, we use the fact that the dominant contribution to the right hand side in (\ref{a2}) are terms with small $b$ and other terms make a negligible contribution.
We also use the equality $\sum_{b=1}^{\infty}1/b^2=\pi ^2/6$.
Finally, by using (\ref{Vc}), $\langle \mu _{0}(\tau )^2 \rangle _c$ satisfies 
\eq{\left \langle \mu _{0}(\tau )^2\right \rangle _c=L\rho \left \langle \nu _{0}(\tau )^2\right \rangle _c=\Theta (L).}
This result shows that the variance of $\mu _{0}(\tau )$ diverges in proportion to $L$ in the thermodynamic limit $L\to \infty$.

Conversely, in the case $J\neq 0$, the right hand side in (\ref{a2}) is
\eqa{\left(\frac{1}{Ml}\right)^2\sum_{b\neq 0}\frac{\lambda _b}{1-\lambda _b}
\simeq \left(\frac{1}{Ml}\right)^2\sum_{b\neq 0}\frac{1+i\frac{2\pi b}{l}\cdot J}{-i\frac{2\pi b}{l}\cdot J+\left(\frac{2\pi b}{l}\right)^2\cdot C}
\simeq \left(\frac{1}{Ml}\right)^2\sum_{b\neq 0}\frac{C-J^2}{J^2}
=\Theta \left(\frac{1}{l}\right)}{Jnonzero}
for large $l$.
Here, we use the fact that $\sum_{b\neq 0}\Im \lambda _b/\left(1-\lambda _b\right)=0$, which is implied by $\Im \lambda _i=-\Im \lambda _{-i}$ and $\mathrm{Re\,}\lambda _i=\mathrm{Re\,}\lambda _{-i}$.
It can be seen from Eq. (\ref{Jnonzero}) that 
\eqa{\left \langle \mu _{0}(\tau )^2\right \rangle _c=L\rho \left \langle \nu _{0}(\tau )^2\right \rangle _c=\Theta (1),}{stresult}
where $\Theta (1)$ indicates that the left hand side in (\ref{stresult}) is independent of $L$.
This result shows that the variance of $\mu _{0}(\tau )$ converges in thermodynamic limit when there is macroscopic flow.

\subsection{Systems in two and more dimensions}

For systems in more than two dimensions, we can perform calculation in a manner similar to that in one dimension.
Therefore, in this subsection, we show an outline of the calculation.

We consider a two-dimensional lattice.
Let $M_x$ and $M_y$ be periods of transition matrix in directions $x$ and $y$, respectively, and $L_x=M_xl_x$ and $L_y=M_yl_y$ be the length of lattice.
$(i, j)\in L_x\times L_y$ represents a position on the lattice.
Let ${\bs v}\in {\mathbb Q}^{L_x\times L_y}$ be the right eigenvectors of transition matrix $S\in \mathbb{R}^{\left(L_x\times L_y\right)^2}$.
In a similar manner to one-dimensional case, it follows from Bloch's theorem that ${\bs v}$ is written as
\eq{\mx{
k_{0,0} & \cdots &k_{M_x-1, 0} & k_{0, 0}e^{i\frac{1}{l_x}\theta _x} &\cdots & k_{M_x-1, 0}e^{i\frac{l_x-1}{l_x}\theta _x} \\
\vdots & \ddots & \vdots & \vdots & \cdots & \vdots \\
k_{0, M_y-1} & \cdots & k_{M_x-1, M_y-1} & k_{0, M_y-1}e^{i\frac{1}{l_x}\theta _x} &\cdots & k_{M_x-1, M_y-1}e^{i\frac{l_x-1}{l_x}\theta _x} \\
k_{0,0}e^{i\frac{1}{l_y}\theta _y} & \cdots &k_{M_x-1, 0}e^{i\frac{1}{l_y}\theta _y} & k_{0, 0}e^{i\frac{1}{l_x}\theta _x}e^{i\frac{1}{l_y}\theta _y} &\cdots & k_{M_x-1, 0}e^{i\frac{l_x-1}{l_x}\theta _x}e^{i\frac{1}{l_y}\theta _y} \\
\vdots & \vdots & \vdots & \vdots & \ddots & \vdots \\
k_{0, M_y-1}e^{i\frac{l_y-1}{l_y}\theta _y} & \cdots & k_{M_x-1, M_y-1}e^{i\frac{l_y-1}{l_y}\theta _y} & k_{0, M_y-1}e^{i\frac{1}{l_x}\theta _x}e^{i\frac{l_y-1}{l_y}\theta _y} &\cdots & k_{M_x-1, M_y-1}e^{i\frac{l_x-1}{l_x}\theta _x}e^{i\frac{l_y-1}{l_y}\theta _y}
},}
where $\theta _x=2\pi b$ ($b=-{l_x}/{2}+1, -{l_x}/{2}+2\cdots {l_x}/{2}$ when $l_x$ is even, and $b=-{(l_x-1)}/{2}, -{(l_x-1)}/{2}+1\cdots {(l_x-1)}/{2}$ when $l_x$ is odd) and $\theta _y=2\pi c$ ($c=-{l_y}/{2}+1, -{l_y}/{2}+2\cdots {l_y}/{2}$ when $l_y$ is even, and $c=-{(l_y-1)}/{2}, -{(l_y-1)}/{2}+1\cdots {(l_y-1)}/{2}$ when $l_y$ is odd).
$k_{0,0}\cdots k_{M_x-1, M_y-1}$ are coefficients.
Here, we write the vector ${\bs v}$ as a matrix only for convenience.
Let $\lambda _{b, c}$ be the corresponding eigenvalue.
Now, we define a transfer matrix $A(h)$ in a manner similar to Eq. (\ref{defA}), and derive $\Lambda _{\mathrm max}(h)$, the maximum eigenvalue of $A(h)$, through a perturbation expansion in powers of $h$.
Eq. (\ref{Vc}) still holds, and Eq. (\ref{a2}) is modified as
\eqa{2a_2\simeq B\left(\frac{1}{M_xM_yl_yl_x}\right)^2\sum_{\substack{{b, c}\\{(b, c)\neq (0, 0)}}}\frac{\lambda _{b, c}}{1-\lambda _{b, c}}.}{2da2}
$\lambda _{b, c}$ is expanded as
\eqa{\lambda _{b, c}=\lambda _{0, 0}+i\left(\frac{2\pi b}{l_x}+\frac{2\pi c}{l_y}\right)\cdot J-\left(\frac{2\pi b}{l_x}+\frac{2\pi c}{l_y}\right)^2\cdot C+\cdots.}{2dlambda}
If $J\neq 0$, we can easily show that $\langle \mu _{\boldsymbol{0}}(\tau )^2\rangle _c$ is independent of $l$, and this implies the absence of the anomaly.
Thus, let us consider the case when $J=0$.
In this case, the second term of the right hand side in (\ref{2dlambda}) is zero.
For simplicity, assume $l_x=l_y=l$. 
By substituting Eq. (\ref{2dlambda}) to Eq. (\ref{2da2}), we obtain 
\begin{align}
B\left(\frac{1}{M_xM_yl^2}\right)^2\sum_{\substack{{b, c}\\{(b, c)\neq (0, 0)}}}\frac{\lambda _{b, c}}{1-\lambda _{b, c}}
&\simeq B\left(\frac{1}{M_xM_yl^2}\right)^2\sum_{\substack{{b, c}\\{(b, c)\neq (0, 0)}}}\frac{l^2}{4\pi ^2\left(b^2+c^2\right)\cdot C} \notag \\
&\simeq B\left(\frac{1}{M_xM_yl^2}\right)^2\sum_{r=1}^{\sqrt{2}l}\frac{l^2}{4\pi ^2r^2\cdot C}2\pi r \notag \\
&=\Theta \left(\frac{1}{l^2}\ln l\right).
\end{align}
Here, we use $\lambda _{0,0}=1$.
We multiply the above equation by a number of particles $M_xM_yl^2\rho$, and obtain 
\eq{\langle \mu _{\boldsymbol{0}}(\tau )^2\rangle _c=\Theta (\ln l).} 
This result means that the variance of $\mu _{\boldsymbol{0}}(\tau )$ shows a logarithmic divergence on $l$ in the thermodynamic limit.

When a system is in three dimensions and there exists no macroscopic flow, we calculate $a_2$ in a similar manner.
The result is
\begin{align}
2a_2
&\simeq B\left(\frac{1}{M_xM_yM_zl^3}\right)^2\sum_{\substack{{b, c, d}\\{(b, c, d)\neq (0, 0, 0)}}}\frac{\lambda _{b, c, d}}{1-\lambda _{b, c, d}} \notag \\
&\simeq B\left(\frac{1}{M_xM_yM_zl^3}\right)^2\sum_{\substack{{b, c, d}\\{(b, c, d)\neq (0, 0, 0)}}}\frac{l^2}{4\pi ^2\left(b^2+c^2+d^2\right)\cdot C} \notag \\
&\simeq B\left(\frac{1}{M_xM_yM_zl^3}\right)^2\sum_{r=1}^{\sqrt{3}l}\frac{l^2}{4\pi ^2r^2\cdot C}4\pi r^2 \notag \\
&=\frac{B}{\left(M_xM_yM_z\right)^2l^3}\frac{\sqrt{3}}{\pi C}. \label{3dmicro}
\end{align}
Multiplying Eq. (\ref{3dmicro}) by $M_xM_yM_zl^3\rho$, we find that  
\eq{\langle \mu _{0}(\tau )^2 \rangle _c=\Theta (1).}
In other words, the variance of $\mu _{\boldsymbol{0}}(\tau )$ converges in the thermodynamic limit.

\

In this section, we calculate the variance of $\mu _{\boldsymbol{0}}(\tau )$.
We can calculate higher-order cumulants in a manner similar to that for variance, and they show the same anomalous system size dependence.
Because the large deviation function and the cumulant generating function are transformed to each other via the Legendre transformation, the above results indicate that $I(\mu )$ shows the anomalous system size dependence.

Note that the ``no macroscopic flow" condition ($J=0$) is wider than the equilibrium condition.
We can easily check that a model which breaks detailed balance but satisfies $J=0$ condition also shows the anomaly.
On more realistic situation, for example, a system that is microscopically irreversible and macroscopically reversible~\cite{reverse} and a system with a non-Gaussian noise~\cite{nongauss} exhibit no macroscopic flow but are out of equilibrium.
Such systems break detailed balance only locally, and the anomaly occurs when detailed balance is broken globally.
Therefore, we expect such systems also shows the anomaly.

\section{Diffusive systems}

In this section, we consider diffusive systems whose probability distribution functions evolve according to the Fokker-Planck equation
\eqa{\frac{\del \mu (\bs{x})}{\del t}+\nabla \cdot \bs{j}(\bs{x})=0}{FPeq}
with current densities
\eqa{{\bs j}(\bs{x}):=\beta D \mu (\bs{x})\cdot ({\bs F}-\nabla U(\bs{x}))-D\nabla \mu (\bs{x}).}{defj}
Here, $D$ is the diffusion coefficient, $U(\bs{x})$ is the potential, $\bs{F}$ is the force, and $\beta$ is the inverse temperature.
Note that random walks and SSEP systems, which are already known to show the anomaly of the local empirical measure (in Ref.~\cite{RW, otSSEP} and the previous section), give diffusion equations similar to Eq. (\ref{FPeq}) in the hydrodynamic limit~\cite{hdl}.
Therefore, analyses in this and the next sections can be applied to such systems.

For such systems, C. Maes {\it et al}~\cite{em} found exact formulas of the large deviation functions for the empirical distribution function defined as $\mu ({\bs x}, \tau ):=\left(\mu _{\bs x}(\tau )\right)_{{\bs x}\in {\mathbb R}^d}$.
Note that $\mu _{\bs x}(\tau )$ is a function of only $\tau$; in contrast, $\mu ({\bs x}, \tau )$ is a function of both $\tau$ and ${\bs x}$.
We can calculate the local empirical measure, $\mu _{\bs 0}(\tau )$, with the empirical distribution function. 
Let $I_{\mathrm DF}(\mu (\bs{x}))$ be the large deviation function for the empirical distribution function $\mu (\bs{x}, \tau )$, namely
\eq{\mathrm{Prob}\left(\mu (\bs{x}, \tau )=\mu (\bs{x})\right)\sim e^{-\tau I_{\mathrm DF}(\mu (\bs{x}))}.}
Hereafter, we abbreviate $\mu (\bs{x})$ to $\mu$.
Ref.~\cite{em} showed that $I_{\mathrm DF}(\mu )$ can be written as
\eqa{I_{\mathrm DF}(\mu)=\frac{\sigma (\mu )-\sigma _V(\mu )}{4},}{ij}
where $\sigma (\mu )$ is the entropy production with $\mu$, $V$ is a modified potential under which $\mu$ is a stationary solution of Eq. (\ref{FPeq}), and $\sigma _V(\mu )$ is the entropy production in the modified system with $V$.
In other words, when $\mu$ is given, $V$ is determined such that
\eqa{\nabla \cdot \bs{j}_V(\bs{x})=\nabla \cdot \left(\beta D\mu (\bs{x})\cdot  (\bs{F}-\nabla V(\bs{x}))-D\nabla \mu (\bs{x})\right)=0}{jvdef}
is satisfied for all $x$.

In this section, we  calculate the local empirical measure by using a contraction principle~\cite{contra}
\eqa{I(\mu _{\bs 0})=\inf_{\mu(\bs{x})} I_{\mathrm DF}(\mu (\bs{x}))}{cont-base}
under constraint conditions $\mu (\bs{0})=\mu _{\bs 0}$ and $\int \mu (\bs{x})d\bs{x}=1$.
Although we can consider systems with any number of dimensions, in order to avoid some technical difficulties, we focus on one-dimensional systems in the argument below.
Henceforth, for simplicity, we set $U=0$.

\subsection{Properties of empirical distribution function}

We consider a one-dimensional periodic boundary system with length $L$, and take a coordinate $-L/2\leq x\leq L/2$.
We use a contraction principle to the origin $x=0$.
Note that the meaning of $L$ in this section is different from that in the previous section.
In the previous section, $L$ is a length before taking the hydrodynamic limit.
In contrast, in this section, $L$ is a length after taking the hydrodynamic limit.
However, in terms of the $L$ dependence, the discussions in this and previous sections give the same results.

The entropy production is written as
\eqa{\sigma (\mu )=\beta \left(\int Fjdx-\frac{d}{dt}\int U\mu dx\right)+\frac{ds(\mu )}{dt}=\int \frac{j^2}{\mu D}dx,}{def-ep}
where $s(\mu )$ denotes the Shannon entropy $s(\mu ):=\int \mu \ln \mu dx$.
Here, the first term in (\ref{def-ep}) represents heat dissipation caused by the force, the second term in (\ref{def-ep}) represents heat dissipation caused by the change of the potential, 
and the third term in (\ref{def-ep}) represents the change of the Shannon entropy.
It follows from Eq. (\ref{defj}) that
\begin{align}
\sigma (\mu )&=L\beta ^2DF^2\bar{\mu}+\int D\frac{\left(\nabla \mu \right)^2}{\mu}dx, \\
\sigma _V(\mu)&=L\beta Fj_V, \label{sigmav}
\end{align}
where $\bar{\mu}$ represents the uniform measure.
Eq. (\ref{defj}) is transformed to
\eqa{j_V\cdot \frac{1}{\mu (x)}=\beta D(F-\nabla V(x))-D\frac{\nabla \mu (x)}{\mu (x)},}{j_V}
where we set $U=V$.
By integrating Eq. (\ref{j_V}) around the ring, $j_V$ satisfies 
\eq{j_V\int_{-L/2}^{L/2}\frac{dx}{\mu}=L\beta D F.}
Set $\delta \mu :=\mu -\bar{\mu}$, and assume that $\delta \mu$ is much smaller than $\bar{\mu}$.
It is implied by Eq. (\ref{ij}) that
\begin{align}
I_{\mathrm DF}(\mu )
&=\frac{1}{4}\left(\int_{-L/2}^{L/2} \frac{D\left(\nabla \mu \right)^2}{\mu}dx+L^2\beta ^2D F^2\left(\frac{1}{\int_{-L/2}^{L/2}\frac{dx}{\bar{\mu}}}-\frac{1}{\int_{-L/2}^{L/2}\frac{dx}{\mu}}\right)\right) \notag \\
&\simeq \frac{1}{4}\left(\int_{-L/2}^{L/2} \frac{D\left(\nabla \mu \right)^2}{\mu}dx+L^2\beta ^2DF^2\frac{\int_{-L/2}^{L/2}\frac{dx}{\mu}-\frac{dx}{\bar{\mu}}}{\left(\int_{-L/2}^{L/2}\frac{dx}{\bar{\mu}}\right)^2}\right) \notag \\
&=\frac{1}{4}\left(\int_{-L/2}^{L/2} \frac{D\left(\nabla \mu \right)^2}{\mu}dx+\beta ^2DF^2\bar{\mu}
^2\int_{-L/2}^{L/2}\frac{1}{\mu}-\frac{1}{\bar{\mu}}dx\right) \notag \\
&=\frac{1}{4}\left(\int_{-L/2}^{L/2} \frac{D\left(\nabla \mu \right)^2}{\mu}dx+\beta ^2DF^2\bar{\mu}
^2\int_{-L/2}^{L/2}\left(\frac{1}{\bar{\mu}}-\frac{\delta \mu}{\bar{\mu}^2}+\frac{\delta \mu ^2}{\bar{\mu}^3}+O\left(\delta \mu ^3\right)\right)-\frac{1}{\bar{\mu}}dx\right) \notag \\
&\simeq \frac{1}{4}\int_{-L/2}^{L/2}\left(\frac{D\left(\nabla \delta \mu \right)^2}{\bar{\mu}}+\frac{\beta ^2DF^2\delta \mu ^2}{\bar{\mu}}\right)dx. \label{IDF}
\end{align}
In the last line, we have used the relation $\int_{-L/2}^{L/2} \delta \mu dx=0$.

\subsection{Contraction}

By using the contraction principle in relation to Eq. (\ref{IDF}), we obtain a minimizing problem
\eqa{I(\mu _0)=\inf_{\delta \mu}\frac{1}{4}\int_{-L/2}^{L/2}\left(\frac{D\left(\nabla \delta \mu \right)^2}{\bar{\mu}}+\frac{\beta ^2DF^2\delta \mu ^2}{\bar{\mu}}\right)dx}{Ist}
with constraint conditions
\begin{align}
\delta \mu (0)&=\delta \mu _0, \label{con-0} \\
\left. \frac{d\delta \mu}{dx}\right|_{x=-L/2}&=\left. \frac{d\delta \mu}{dx}\right|_{x=L/2},\label{con-L} \\
\int_{-L/2}^{L/2}\delta \mu (x)dx&=0, \label{concv}
\end{align}
where we set $\delta \mu _0:=\mu _0-\bar{\mu}$.
Let $\delta \mu _{\mathrm s}(x)$ be a function that minimizes the right hand side in (\ref{Ist}).
By noting the constraint condition (\ref{concv}), we use the variational method and obtain a differential equation
\eqa{\nabla ^2\delta \mu _{\mathrm s}(x)-\beta ^2F^2\delta \mu _{\mathrm s}(x)=\mathrm{const}.}{1dmin}

Eq. (\ref{1dmin}) has qualitatively different solutions depending on whether $F=0$ or $F\neq 0$.
When $F=0$（no flow case), the solution to Eq. (\ref{1dmin}) is derived as
\eqa{\delta \mu _{\mathrm s}(x)=
\begin{cases}
\dis \frac{6\delta \mu _0}{L^2}\left(x-\frac{L}{2}\right)^2-\frac{\delta \mu _0}{2} &\mathrm{for }\ 0\leq x\leq \frac{L}{2}, \\
\\
\dis \frac{6\delta \mu _0}{L^2}\left(x+\frac{L}{2}\right)^2-\frac{\delta \mu _0}{2} &\mathrm{for }\ -\frac{L}{2}\leq x\leq 0.
\end{cases}}{1kai}
Substituting Eq. (\ref{1kai}) into Eq. (\ref{Ist}), $I(\mu _0)$ satisfies
\eqa{I(\mu _0)=\frac{D}{4\bar{\mu}}\left(\frac{12(\mu _0-\bar{\mu})}{L^2}\right)^2\frac{L^3}{12}=\Theta \left( \frac{1}{L}\right).}{Ikai}
This result shows anomalous dependence on $L$.
With respect to variance, Eq. (\ref{Ikai}) indicates that the variance of $\mu _0$ is proportion to $L$, and this is consistent with results in Sec. I\hspace{-.1em}I.

In contrast, when $F\neq 0$ (flow case), the solution to Eq. (\ref{1dmin}) is expressed as
\eqa{\delta \mu _{\mathrm s}=
\begin{cases}
A+Be^{-Sx}+Ce^{Sx} &\mathrm{for}\ 0\leq x\leq \frac{L}{2}, \\
A+Be^{Sx}+Ce^{-Sx}  &\mathrm{for}\ -\frac{L}{2}\leq x\leq 0,
\end{cases}
}{kaist}
where $S=|\beta F|$.
The constraint conditions (\ref{con-0}), (\ref{con-L}), and (\ref{concv}) determine $A$, $B$, and $C$ as
\begin{align}
A&=\frac{\left(e^{-LS}-1\right)\delta \mu _0}{1+LS/2+\left(LS/2-1\right)e^{-LS}}, \\
B&=\frac{\delta \mu _0}{1+\frac{2}{LS}+\left(1-\frac{2}{LS}\right)e^{-LS}}, \\
C&=\frac{e^{-LS}\delta \mu _0}{1+\frac{2}{LS}+\left(1-\frac{2}{LS}\right)e^{-LS}}.
\end{align}
Substituting Eq. (\ref{kaist}) into Eq. (\ref{Ist}), we find that 
\eq{I(\mu _0)=\Theta (1).}
This result shows that there exists no anomalous dependence on $L$ when flow exists in the system.

\section{Phenomenological arguments}

In the previous sections, it is suggested that the necessary and sufficient conditions for this anomaly are i) there is no macroscopic flow, and ii) the dimension of systems is one or two. 
In this section, we derive these conditions from a phenomenological viewpoint.

We consider fluid systems that show the long-time tail behavior of density.
The long-time tail is a phenomenon in which time correlation functions of $\Delta \rho$, a fluctuation of density, decay not exponentially but as power law.
This phenomenon was first found in simulations of hard spheres as a property of velocity autocorrelation functions~\cite{Alder}.
Thereafter, it has been found that many conserved quantities in various fluid systems show such phenomena~\cite{ltt, ltt2, ltt3, ssltt, lttstress}.

We assume the specific form of the long-time tail behavior as follows.
Let $\Delta \rho _{\boldsymbol{0}}(t)$ be a fluctuation of density at the origin ${\bs x}={\boldsymbol{0}}$ and at time $t$.
We consider a $d$-dimensional cubic system with sides of $L$ and suppose that there exists no flow. 
In this setting, we assume that the density autocorrelation function satisfies
\begin{align}
\la \Delta \rho _{\boldsymbol{0}}(0)\Delta \rho _{\boldsymbol{0}}(t)\ra 
\begin{cases}
=f(t) &\text{: $0\leq t<C$}, \\
\propto t^{-d/2} &\text{: $C\leq t<g(L)$}, \\
\propto c^{-t} &\text{: $g(L)\leq t$},
\end{cases}
\label{ltt}
\end{align}
asymptotically in large $t$.
Here, $C$ denotes the time when the long-time tail behavior starts to appear, and $C$ is independent of $L$.
$f(t)$ represents the behavior of autocorrelation functions until $t=C$.
$g(L)$ satisfies $g(L)\propto L^2$.
This relation is true if the system shows normal diffusion. 
If the system size is infinite, $\la \Delta \rho _{\boldsymbol{0}}(0)\Delta \rho _{\boldsymbol{0}}(t)\ra \propto t^{-d/2}$ for $C<t$.
The symbol ``$\lesssim t^{-1-d/2}$" indicates that a speed of convergence to $0$ under $t\to \infty$ is the same as or faster than $t^{-1-d/2}$.

Conversely, in the case that the system size is infinite and there is macroscopic flow, we assume the relation
\begin{align}
\la \Delta \rho _{\boldsymbol{0}}(0)\Delta \rho _{\boldsymbol{0}}(t)\ra 
\begin{cases}
=f(t) &\text{: $0\leq t<C$}, \\
\propto t^{-d/2} &\text{: $C\leq t<h(F)$}, \\
\lesssim t^{-1-d/2} &\text{: $h(F)\leq t$}. 
\end{cases}
\label{sltt}
\end{align}
Here, $F$ is the force that causes the macroscopic flow.
$h(F)$ is a function of $F$, and $h(F)$ satisfies $h(F)\propto F^{-2}$.
The symbol ``$\lesssim t^{-1-d/2}$" indicates that a speed of convergence to $0$ under $t\to \infty$ is the same as or faster than $t^{-1-d/2}$.
Note that the formula of $f(t)$ and the value of $C$ are different from Eq. (\ref{ltt}).
Systems considered in Sec. I\hspace{-.1em}I\hspace{-.1em}I show such long-time tail behavior (\ref{ltt}) and (\ref{sltt}) (see Appendix).

A mathematical structure of the anomaly of the local empirical measure is very similar to that of the long-time tail.
Owing to this, in this section, with the assumption of the long-time tail (\ref{ltt}) and (\ref{sltt}), we derive the anomalous dependence of $\langle \mu _{0}(\tau )^2 \rangle _c$ on both $\tau$ and $L$.

\subsection{Derivation of anomalous time dependence}

Suppose that there exists no flow in a system.
First, we take the thermodynamic limit $L\to \infty$, and then, we take a large $\tau$.
By using Eq. (\ref{ltt}), we calculate the dependence of $\langle \mu _{0}(\tau )^2 \rangle _c$ on $\tau$ as
\begin{align}
\langle \mu _{0}(\tau )^2 \rangle _c
&=\la \frac{1}{{\tau}^2}\int_0^{\tau} \int_0^{\tau} dt'dt''\Delta \rho _{\boldsymbol{0}}(t') \Delta \rho _{\boldsymbol{0}}(t'') \ra \notag \\
&=\frac{1}{{\tau}^2}\cdot 2\int_0^{\tau} dt'''({\tau}-t''')\la \Delta \rho _{\boldsymbol{0}}(0) \Delta \rho _{\boldsymbol{0}}(t''') \ra  \notag \\
&\simeq \frac{1}{{\tau}^2}\cdot 2\left({\tau}\int_0^Cdt'''f(t''')+\int_C^{\tau}dt'''({\tau}-t''')\cdot at'''^{-d/2}\right) \notag \\
&\simeq
\begin{cases} 
\frac{1}{{\tau}^2}\cdot 2\left({\tau}K+a'{\tau}^{2-d/2}\right) &\mathrm{for}\ d\neq 2,\\
\frac{1}{{\tau}^2}\cdot 2\left({\tau}K+a'{\tau}\ln {\tau}\right) &\mathrm{for}\ d=2, 
\end{cases} \label{Tsin}
\end{align}
where $K=\int_0^Cdtf(t)$.
$a$ and $a'$ are constants.
In the third line, we have used an approximation $\tau -C\simeq \tau$.

It follows from Eq. (\ref{Tsin}) that the dependence of $\langle \mu _{0}(\tau )^2 \rangle _c$ on ${\tau}$ is
\eqa{\langle \mu _{0}(\tau )^2 \rangle _c \propto
\begin{cases}
\dis \frac{1}{\sqrt{{\tau}}} &\mathrm{for}\ d=1, \\
\dis \frac{\ln {\tau}}{{\tau}} &\mathrm{for}\ d=2, \\
\dis \frac{1}{{\tau}} &\mathrm{for}\ d\geq 3.
\end{cases}
}{Tre}
This result represents the anomalous dependence on ${\tau}$ in one and two dimensions.

\subsection{Derivation of anomalous system size dependence}

Suppose that there exists no flow in a system.
First, we take the long-time limit ${\tau}\to \infty$, and then, we take a large $L$.
By using Eq. (\ref{ltt}), we calculate the dependence of $\langle \mu _{0}(\tau )^2 \rangle _c$ on $L$ as
\begin{align}
\lim_{{\tau}\to \infty}\langle \mu _{0}(\tau )^2 \rangle _c 
&=\lim_{{\tau}\to \infty}\frac{1}{{\tau}^2}\cdot 2\int_0^{\tau} dt'''({\tau}-t''')\la \Delta \rho _{\boldsymbol{0}}(0) \Delta \rho _{\boldsymbol{0}}(t''') \ra \notag \\
&\simeq \lim_{{\tau}\to \infty}\frac{1}{{\tau}^2}\cdot 2\left({\tau}\int_0^Cdt'''f(t''')+\int_C^{g(L)}dt'''({\tau}-t''')\cdot at'''^{-d/2}\right) \notag \\
&\simeq
\begin{cases} 
\dis \lim_{{\tau}\to \infty}\frac{1}{{\tau}^2}\cdot 2\left({\tau}K+a'{\tau}\left(g(L)^{1-d/2}-C^{1-d/2}\right)\right) &\mathrm{for}\ d\neq 2,\\
\dis \lim_{{\tau}\to \infty}\frac{1}{{\tau}^2}\cdot 2\left({\tau}K+a'{\tau}\ln \frac{g(L)}{C}\right) &\mathrm{for}\ d=2, 
\end{cases} \label{Vsin}
\end{align}
where $a$ and $a'$ are constants.
It follows from Eq. (\ref{Vsin}) that the dependence of $\langle \mu _{0}(\tau )^2 \rangle _c$ on $L$ is
\eqa{\lim_{{\tau}\to \infty}\langle \mu _{0}(\tau )^2 \rangle _c \propto
\begin{cases}
\dis L &\mathrm{for}\ d=1, \\
\dis \ln L &\mathrm{for}\ d=2, \\
\dis \mathrm{const} &\mathrm{for}\ d\geq 3,
\end{cases}
}{Vre}
for large $L$.
This result represents the anomalous dependence on $L$ in one and two dimensions.

\subsection{No anomaly with macroscopic flow}

Suppose that macroscopic flow exists in a system.
First, we take the thermodynamic limit $L\to \infty$, and then, we take large $\tau$.
From Eq. (\ref{ltt}), the dependence of $\langle \mu _{0}(\tau )^2 \rangle _c$ on $\tau$ is calculated as
\begin{align}
\langle \mu _{0}(\tau )^2 \rangle _c 
&=\la \frac{1}{{\tau}^2}\int_0^{\tau} \int_0^{\tau} dt'dt''\Delta \rho _{\boldsymbol{0}}(t') \Delta \rho _{\boldsymbol{0}}(t'') \ra \notag \\ 
&=\frac{1}{{\tau}^2}\cdot 2\int_0^{\tau} dt'''({\tau}-t''')\la \Delta \rho _{\boldsymbol{0}}(0) \Delta \rho _{\boldsymbol{0}}(t''') \ra \notag \\
&\simeq \frac{1}{{\tau}^2}\cdot 2\left({\tau}\int_0^Cdt'''f(t''')+\tau \int_C^{h(F)}dt'''bt'''^{-d/2}+ \int_{h(F)}^{\tau}dt'''({\tau}-t''')\cdot at'''^{-1-d/2}\right) \notag \\
&\simeq
\frac{1}{{\tau}^2}\cdot 2\left({\tau}K+a'{\tau}^{1-d/2}\right), \label{ssltt}
\end{align}
where $a$, $b$, and $a'$ are constants.
As implied by $d\geq 1$, Eq. (\ref{ssltt}) shows standard dependence on $\tau$; 
\eq{\langle \mu _{0}(\tau )^2 \rangle _c \propto \frac{1}{{\tau}}.}
This result shows that the anomaly vanishes when flow exists in a system.

\section{Concluding remarks}

The essential conditions for the anomaly in systems with translational symmetry are i) there exists no macroscopic flow, and ii) the dimension of the system is one or two.
When these conditions are satisfied, a power law behavior of statistical properties is observed.
Concretely, in Sec. I\hspace{-.1em}I\hspace{-.1em}I, we show that $\mu _{\mathrm s}(x)$, the function minimizing $I(\mu (x))$, decays from the origin $x=0$ as a power law in the space direction.
In Sec. I\hspace{-.1em}V, we show that the time correlation function of $\Delta \rho _{\boldsymbol{0}}(t)$ has a power law tail structure in the time direction, and the anomaly is derived directly from this structure.
Hence, this anomaly can be seen not only for the local empirical measure but also for other time-averaged conserved quantities that show the long-time tail behavior, such as the velocity~\cite{Alder, ltt, ltt2, ltt3, ssltt} and the stress tensor~\cite{lttstress}.

In particular, for systems in which total momentum is conserved, we expect that variance for time-averaged pressure shows anomalous time and system size dependence even in three dimensions.
Concretely, when we take the long-time limit first, the variance behaves as $\Theta (L)$, and when we take the thermodynamic limit, the variance behaves as $\Theta (1/\sqrt{\tau})$ for any dimensions.
The reason why the anomaly remains even in three and more dimensions is as follows; for a system in $d$-dimensions, a side wall is in $d-1$-dimensions.
The difference between the dimensions of systems and side walls is always one, and thus, with projections, the variance for time-averaged pressure seems to show the same anomaly as that for the one-dimensional case. 
This is a future problem.

In conclusion, we have two comments.
First, in Sec. I\hspace{-.1em}I\hspace{-.1em}I we show that the anomaly is caused by the contraction process.
Here, it is known that some nonequilibrium steady systems show anomalous fluctuations called long-range correlations~\cite{lrc}.
When this anomaly occurs, a fluctuation decays not exponentially but as power law in the space direction.
Recent studies show that the long-range correlation can be understood as the anomaly caused by a contraction process for the time direction, and the derivation also uses a modified potential~\cite{Bertini}.
This is very similar to the discussion in Sec. I\hspace{-.1em}I\hspace{-.1em}I, and therefore the long-range correlation and the anomaly we discuss in this paper seem to have a common origin.
This is also a future problem.

Second, in recent years, new universal relations in nonequilibrium thermodynamics have been investigated~\cite{HatanoSasa, KNST}.
With regard to this development, Ref.~\cite{eC} has proposed a novel type of inequality in nonequilibrium thermodynamics, in the sense that it is written with the large deviation function for the empirical distribution function.
It would be remarkable if the anomaly reported in this paper is related to nonequilibrium thermodynamics.

\begin{acknowledgements}
The author thanks S.-I. Sasa for providing useful advices and fruitful discussions. The author also thanks M. Otsuki and T. Nemoto for providing helpful comments.

\end{acknowledgements}

\section*{Appendix: Proof of the long-time tail behavior}
\appnum{A}

In this appendix, we prove the long-time tail behavior (\ref{ltt}) and (\ref{sltt}) for diffusive systems whose distribution functions evolve according to the Fokker-Planck equation (\ref{FPeq}) with $U=0$.
Suppose that the system is in $d$ dimensions under periodic boundary conditions with length $L$. 
We also assume that the initial condition is $\Delta \rho _{\boldsymbol x}(0)=0$ for ${\boldsymbol x}\neq 0$.

First, for $F=0$, we derive Eq. (\ref{ltt}).
The evolution of $\Delta \rho$ is described by
\eqa{\frac{\del \left(\Delta \rho \right)}{\del t}=D\nabla ^2\left(\Delta \rho \right).}{F0FP}
From the Fourier transforms of Eq. (\ref{F0FP}), it is shown that
\eqa{\Delta \rho _{\boldsymbol 0}(t)=\sum_{\bs k}e^{-D\left|{\bs k}\right|^2t},}{0Fourier}
where ${\bs k}$ takes ${\bs k}={2\pi}/{L}\cdot {\bs j}$ for all ${\boldsymbol j}\in{\mathbb N}^d$ except ${\bs j}={\bs 0}$.
When ${4D\pi ^2t}/{L^2}\ll 1$, we can transform Eq. (\ref{0Fourier}) to an integral form as
\eqa{\Delta \rho _{\boldsymbol 0}(t)=\int e^{-D\left|{\bs k}\right|^2t}d{\bs k}=t^{-\frac{d}{2}}\int e^{-D\left|{\bs k'}\right|^2}d{\bs k'\propto t^{-d/2}}.}{0int}
Here, we set $\sqrt{t}{\bs k}={\bs k'}$.
In contrast, when ${4D\pi ^2t}/{L^2}\gg 1$, the right hand side in (\ref{0Fourier}) is written as
\eqa{\sum_{\bs k}e^{-D\left|{\bs k}\right|^2t}\simeq \sum_{|{\bs k}|=2\pi /L}e^{-D\left|{\bs k}\right|^2t}+\sum_{n=2}^{\infty}\sum_{|{\bs k}|=2\pi \sqrt{n}/L}e^{-D\left|{\bs k}\right|^2t}\simeq 2d\cdot e^{-D\left|\frac{2\pi}{L}\right|^2t}\propto c^{-t}.}{0sum}
Here, we use
\eq{\sum_{n=2}^{\infty}\sum_{|{\bs k}|=2\pi \sqrt{n}/L}e^{-D\left|{\bs k}\right|^2t}
\leq \sum_{n=2}^{\infty}(2d)^ne^{-\frac{4D\pi ^2nt}{L}}=\sum_{n=2}^{\infty}e^{\left(-\frac{4D\pi ^2t}{L}+\ln 2d\right)n}=o\left(e^{-D\left|\frac{2\pi}{L}\right|^2t}\right)}
Eqs. (\ref{0int}) and (\ref{0sum}) satisfy the long-time tail behavior (\ref{ltt}).

Next, for infinite size systems with $F\neq 0$, we derive Eq. (\ref{sltt}).
The evolution of $\Delta \rho$ is described by
\eqa{\frac{\del \left(\Delta \rho \right)}{\del t}=-\beta D{\bs F}\cdot \nabla \left(\Delta \rho \right)+D\nabla ^2\left(\Delta \rho \right).}{Fnon0FP}
From the Fourier transforms of Eq. (\ref{Fnon0FP}), it is shown that
\eqa{\Delta \rho _{\boldsymbol 0}(t)=\int e^{\left(-D\left|{\bs k}\right|^2-i\beta D\bs{F}\cdot \bs{k}\right)t}d{\bs k}=\int e^{-D\left|{\bs k}\right|^2t}\cos (\beta D\bs{F}\cdot \bs{k}t)d{\bs k}.}{non0int}
When $\beta D\left|\bs{F}\right|t \ll \sqrt{Dt}$, we perform calculation in a manner similar to Eq. (\ref{0int}) as
\eqa{\Delta \rho _{\boldsymbol 0}(t)=\int e^{-D\left|{\bs k}\right|^2t}\cos (\beta D\bs{F}\cdot \bs{k}t)d{\bs k}\simeq \int e^{-D\left|{\bs k}\right|^2t}d{\bs k}=t^{-\frac{d}{2}}\int e^{-D\left|{\bs k'}\right|^2}d{\bs k'}\propto t^{-d/2}.}{non0short}
In contrast, when $\beta D\left| \bs{F}\right|t \gg \sqrt{Dt}$, we use an approximation for large $A$
\eqa{\int_{k_1}^{k_2}f(k)\cos (2\pi Ak)dk\simeq \frac{1}{4A^2}\int_{k_1}^{k_2}\frac{d^2f}{dk^2}dk,}{non0app}
which follows from
\eq{\int_k'^{k'+\frac{1}{A}}f(k)\cos (2\pi Ak)dk\simeq \frac{1}{4A^3}\left.\frac{d^2f}{dk^2}\right|_{k=k'}.}
We divide the vector $\bs{k}$ into $\bs{k}_{\perp}$ and $\bs{k}_{\parallel}$ ($\bs{k}=\bs{k}_{\perp}+\bs{k}_{\parallel}$), where $\bs{k}_{\perp}$ is perpendicular to $\bs{F}$, and $\bs{k}_{\parallel}$ is parallel to $\bs{F}$.
By using the approximation (\ref{non0app}), we obtain
\begin{align}
\int e^{-D\left|{\bs k}\right|^2t}\cos (\beta D\bs{F}\cdot \bs{k}t)d{\bs k}
&=\int \int e^{-D\left(\left|{\bs k}_{\perp}\right|^2+k_{\parallel}^2\right)t}\cos (\beta D\left|F\right|k_{\parallel}t)d{\bs k}_{\perp}dk_{\parallel} \notag \\
&\simeq t^{-\frac{d-1}{2}}\int e^{-D\left|{\bs k'}_{\perp}\right|^2}d{\bs k'}_{\perp}\cdot \left(\frac{\pi}{\beta FDt}\right)^2\int \frac{d^2}{dk^2}\left(e^{-Dk_{\parallel}^2t}\right)dk_{\parallel} \notag \\
&=C\cdot t^{-\frac{d-1}{2}}\cdot \frac{1}{t}\int \left(-D+D^2k_{\parallel}^2t\right)e^{-Dk_{\parallel}^2t}dk_{\parallel} \notag \\
&\lesssim t^{-1-d/2}.\label{non0long}
\end{align}
Here, $C$ is a constant.
Eqs. (\ref{non0short}) and (\ref{non0long}) satisfy the long-time tail behavior (\ref{sltt}).

\end{document}